\documentclass[12pt,a4paper]{article}
\usepackage{graphics}
\usepackage{amsfonts}
\usepackage{amssymb}

\def\hybrid{\topmargin -20pt    \oddsidemargin 0pt
        \headheight 0pt \headsep 0pt
        \textwidth 6.25in       % A4 paper
        \textheight 9.5in       % A4 paper
        \marginparwidth .875in
        \parskip 5pt plus 1pt   \jot = 1.5ex}

\hybrid

\def\baselinestretch{1.2}

\catcode`\@=11

\def\marginnote#1{}
\newcount\hour
\newcount\minute
\newtoks\amorpm
\hour=\time\divide\hour by60
\minute=\time{\multiply\hour by60 \global\advance\minute by-\hour}
\edef\standardtime{{\ifnum\hour<12 \global\amorpm={am}%
        \else\global\amorpm={pm}\advance\hour by-12 \fi
        \ifnum\hour=0 \hour=12 \fi
        \number\hour:\ifnum\minute<10 0\fi\number\minute\the\amorpm}}
\edef\militarytime{\number\hour:\ifnum\minute<10 0\fi\number\minute}
\def\draftlabel#1{{\@bsphack\if@filesw {\let\thepage\relax
   \xdef\@gtempa{\write\@auxout{\string
      \newlabel{#1}{{\@currentlabel}{\thepage}}}}}\@gtempa
   \if@nobreak \ifvmode\nobreak\fi\fi\fi\@esphack}
        \gdef\@eqnlabel{#1}}
\def\@eqnlabel{}
\def\@vacuum{}
\def\draftmarginnote#1{\marginpar{\raggedright\scriptsize\tt#1}}

\def\draft{\oddsidemargin -.5truein
        \def\@oddfoot{\sl preliminary draft \hfil
        \rm\thepage\hfil\sl\today\quad\militarytime}
        \let\@evenfoot\@oddfoot \overfullrule 3pt
        \let\label=\draftlabel
        \let\marginnote=\draftmarginnote
   \def\@eqnnum{(\theequation)\rlap{\kern\marginparsep\tt\@eqnlabel}%
\global\let\@eqnlabel\@vacuum}  }

\def\preprint{\twocolumn\sloppy\flushbottom\parindent 2em
        \leftmargini 2em\leftmarginv .5em\leftmarginvi .5em
        \oddsidemargin -.5in    \evensidemargin -.5in
        \columnsep .4in \footheight 0pt
        \textwidth 10.in        \topmargin  -.4in
        \headheight 12pt \topskip .4in
        \textheight 6.9in \footskip 0pt
        \def\@oddhead{\thepage\hfil\addtocounter{page}{1}\thepage}
        \let\@evenhead\@oddhead \def\@oddfoot{} \def\@evenfoot{} }

\def\numberbysection{\@addtoreset{equation}{section}
        \def\theequation{\thesection.\arabic{equation}}}

\def\underline#1{\relax\ifmmode\@@underline#1\else
        $\@@underline{\hbox{#1}}$\relax\fi}

\def\titlepage{\@restonecolfalse\if@twocolumn\@restonecoltrue\onecolumn
     \else \newpage \fi \thispagestyle{empty}\c@page\z@
        \def\thefootnote{\fnsymbol{footnote}} }

\def\endtitlepage{\if@restonecol\twocolumn \else \newpage \fi
        \def\thefootnote{\arabic{footnote}}
        \setcounter{footnote}{0}}  %\c@footnote\z@ }

\catcode`@=12
\relax

\def\figcap{\section*{Figure Captions\markboth
        {FIGURECAPTIONS}{FIGURECAPTIONS}}\list
        {Figure \arabic{enumi}:\hfill}{\settowidth\labelwidth{Figure
999:}
        \leftmargin\labelwidth
        \advance\leftmargin\labelsep\usecounter{enumi}}}
 \relax
\def\tablecap{\section*{Table Captions\markboth
        {TABLECAPTIONS}{TABLECAPTIONS}}\list
        {Table \arabic{enumi}:\hfill}{\settowidth\labelwidth{Table
999:}
        \leftmargin\labelwidth
        \advance\leftmargin\labelsep\usecounter{enumi}}}
 \relax
\def\reflist{\section*{References\markboth
        {REFLIST}{REFLIST}}\list
        {[\arabic{enumi}]\hfill}{\settowidth\labelwidth{[999]}
        \leftmargin\labelwidth
        \advance\leftmargin\labelsep\usecounter{enumi}}}
 \relax
\makeatletter
\newcounter{pubctr}
\def\publist{\@ifnextchar[{\@publist}{\@@publist}}
\def\@publist[#1]{\list
        {[\arabic{pubctr}]\hfill}{\settowidth\labelwidth{[999]}
        \leftmargin\labelwidth
        \advance\leftmargin\labelsep
        \@nmbrlisttrue\def\@listctr{pubctr}
        \setcounter{pubctr}{#1}\addtocounter{pubctr}{-1}}}
\def\@@publist{\list
        {[\arabic{pubctr}]\hfill}{\settowidth\labelwidth{[999]}
        \leftmargin\labelwidth
        \advance\leftmargin\labelsep
        \@nmbrlisttrue\def\@listctr{pubctr}}}
 \relax
\makeatother
\newskip\humongous \humongous=0pt plus 1000pt minus 1000pt

\newif\ifdtup

\relax

\def\be{\begin{equation}}
\def\ee{\end{equation}}
\def\ba{\begin{eqnarray}}
\def\ea{\end{eqnarray}}

\def\del{\partial}

\def\a{\alpha}

\def\d{\delta}

\def\th{\theta}

\def\m{\mu}
\def\n{\nu}

\def\Om{\Omega}
\def\l{\lambda}

\def\no{\noindent}

\def\qq{\qquad}

\def\IR{\relax{\rm I\kern-.18em R}}

\def \ha {{1\over 2}}

\def \ov {\over}

\def\IR{\relax{\rm I\kern-.18em R}}
\def\inv{^{\raise.15ex\hbox{${\scriptscriptstyle -}$}\kern-.05em 1}}

\begin{document}
%\draft

\renewcommand{\theequation}{\arabic{equation}}

\newcommand{\beq}{\begin{equation}}
\newcommand{\eeq}[1]{\label{#1}\end{equation}}
\newcommand{\ber}{\begin{eqnarray}}
\newcommand{\eer}[1]{\label{#1}\end{eqnarray}}
\newcommand{\eqn}[1]{(\ref{#1})}
\begin{titlepage}
\begin{center}

\hfill CERN-TH/99-248\\
\hfill hep--th/9908116\\

\vskip .8in

{\large \bf Non-standard compactifications with mass gaps \\ and Newton's law}

\vskip 0.6in

{\bf A. Brandhuber }\phantom{x}and\phantom{x} {\bf K. Sfetsos}
\vskip 0.1in
{\em Theory Division, CERN\\
     CH-1211 Geneva 23, Switzerland\\
{\tt brandhu,sfetsos@mail.cern.ch}}\\
\vskip .2in

\end{center}

\vskip .5in

\centerline{\bf Abstract}

\no
The four-dimensional Minkowski 
space-time is considered as a three-brane embedded 
in five dimensions, using solutions of five-dimensional supergravity.
These backgrounds have a string theoretical interpretation 
in terms of D3-brane distributions. 
By studying linear fluctuations of the graviton we find 
a zero-mode representing the massless graviton in four-dimensional
space-time. 
The novelty of our models is that the graviton spectrum has 
a genuine mass gap (independent of the position of the world-brane)
above the zero-mode or it is discrete.
Hence, an effective four-dimensional theory on a brane that 
includes the massless graviton mode is well defined.
The gravitational force between point particles deviates from the 
Newton law by Yukawa-type corrections, which we compute explicitly.
We show that the parameters of our solutions can be chosen such that 
these corrections lie within experimental bounds.

\vskip 3 cm
\noindent
CERN-TH/99-248\\
August 1999\\
\end{titlepage}
\vfill
\eject

\def\baselinestretch{1.2}
\baselineskip 16 pt
\noindent

\section{Introduction}

The idea that our four-dimensional Minkowski space-time $M_4$ can be viewed
as a three-brane embedded in some higher-dimensional curved space-time 
is appealing and attracted attention already some years ago \cite{GW}.
It was recently revived in several works. In particular in 
\cite{rasu1,rasu2} 
an alternative to the resolution of the mass hierarchy problem 
between the Planck ($M_{\rm Pl}\sim 10^{19}$ GeV) and electroweak 
($M_{\rm E} \sim 1$ TeV) scales
using extra large dimensions \cite{AHDD} was presented. 
Indeed, in the latter scenario
a compactification on a general compact manifold $B_n$ with $n\ge 2$ 
requires that the size of a typical extra dimension 
be of order $R=1$ mm or smaller, 
which is in principle accessible to experiments in the near future 
(see, for instance, \cite{L} and references therein).
However, as far as the hierarchy problem is concerned,
the original mass hierarchy of ${M_{\rm Pl}\ov M_{\rm E}}\sim 10^{16}$ 
is replaced by a new one
of the same order of magnitude between the electroweak scale and the size of
the large dimensions. 
In \cite{rasu1,rasu2} a model based on a slice of the five-dimensional 
anti-de Sitter space ($AdS_5$) was proposed.
The effective four-dimensional Planck constant was determined by the 
curvature of the embedding space rather 
than the size of the extra dimension, which is of the same order
as the fundamental five-dimensional scale. 
Assuming that the square of the fundamental 
five-dimensional mass scale and the curvature of $AdS_5$ are of the 
same order of $M_{\rm Pl}$, one shows that the 
required hierarchy is of the order of $10^2$.

The scenario of \cite{rasu1,rasu2} has the desired feature that there 
exists a square-normalizable state representing a massless graviton. 
However, there is in addition a continuum of massive 
modes with no mass gap separating them from the massless one. 
If the curvature of $AdS_5$ is 
of the order of the four-dimensional effective Planck scale, then 
this gives no measurable effect to modifications of, for instance, the
Newton law \cite{rasu2}. 
However, given our present-day experimental data, 
this might not be the case and such corrections may exist \cite{L}.
Moreover, in order to have a well-defined effective field theory, 
it is desirable to have solutions where the massless graviton is separated 
from the other massive modes by a mass gap. 
These massive modes can be continuous or discrete, 
with no immediate effect in phenomenological considerations. 
It is the purpose of this paper to construct such models.

It is also desirable to have a string theoretical construction for our models.
The model in \cite{rasu1,rasu2} can be thought of as a compactification 
of the near-horizon limit of the solution for a large number of 
coinciding D3-branes in type-IIB string theory on 
$S^5$, and a subsequent truncation of the range of the extra fifth dimension. 
The background in \cite{rasu2} with one three-brane shares many 
features with domain wall solutions of four-dimensional supergravity 
theories, which were studied extensively in the literature \cite{cvetic}. 
A geometry similar to that of the set-up in \cite{rasu1} appears in strongly 
coupled heterotic string theories, which arise in Calabi--Yau 
compactifications of the Horava--Witten model \cite{hw}. 
In \cite{stelle} a domain wall
solution with two three-branes at the two boundaries was constructed, 
which is suitable for a further reduction to four-dimensional supergravity.
In \cite{verlinde} a ten-dimensional background of a configuration of
D3-branes in an orientifold of type-II string theory 
was presented, which interpolates between $AdS_5 \times S^5$ and
$M_4 \times T^6$ and contains four-dimensional
gravity on the branes. 
It is closely related to the set-up in \cite{rasu1,rasu2}
and can be viewed as an extension of the AdS/CFT correspondence \cite{malda}
to  boundary theories including gravity. It is obvious \cite{kehagias}
that the scenario of \cite{rasu1,rasu2} applies to all minima of 
the potential of the five-dimensional gauged supergravity \cite{gausug},
supersymmetric or not,
although the latter minima may not be stable (at least perturbatively).

In this paper  we consider backgrounds of  
continuous distributions \cite{KLT,Sfe1} of D3-branes of type-IIB string 
theory in the near horizon limit, which preserve sixteen supersymmetries. 
We will consider in detail two examples:
one that represents D3-branes uniformly distributed
over a disc and one where the distribution is over a three-sphere.
These examples were studied before in \cite{FGPW,BS} in connection with
the Coulomb branch of strongly coupled $N=4$ super Yang-Mills
theories within the AdS/CFT correspondence. They possess many of the desired 
features we would like to have, 
in particular they have a mass gap or a discrete spectrum
for the case of the disc and sphere distributions, respectively.
However, the massless mode is not square normalizable and this is is one
issue we address and solve in this paper.
Although we will concentrate on these two backgrounds for concreteness, 
we believe that one can in this context study a much larger
class of distributions of D3-branes e.g.
shell-type distributions which were also studied before
in \cite{CheGid} as models for the Coulomb  branch of $N=4$ super Yang-Mills.

The outline of the paper is as follows: in section 2 
we derive the five-dimensional backgrounds in the presence
of a three-brane that cuts off the space in the fifth direction.
In section 3 we study the spectrum of linear fluctuations of the graviton 
in these backgrounds. 
It turns out that the equation for gravitons polarized 
in the directions parallel to the brane is identical to 
the massless scalar equation in the same backgrounds.
We will show that a normalizable zero-mode solution, 
representing the massless graviton, exists, whereas the massive modes are
separated by a mass gap. 
We use these models in section 4 to study possible measurable effects 
manifested as corrections to Newton's law, which 
turn out to be of Yukawa-type. Our analysis has certain similarities with
that performed for the case of compactifications on general 
compact manifolds in \cite{kesf}. The range of the exponential correction is 
associated with the wavelength of the lightest massive state, whereas its
strength is related to the value of the corresponding wave function at 
the position of the brane. 
In section 5 we present our conclusions and some
future directions of this work.
We have also written an appendix, containing some details of the relation
between ten- and five-dimensional backgrounds and the 
corresponding five-dimensional gauged supergravity theories.

\section{The models}

In the following we exploit the fact that there exists a big class
of ten-dimensional backgrounds, besides $AdS_5 \times S^5$,
that can be reduced to five-dimensional models.
After the reduction to five dimensions the backgrounds
become warped products of four-dimensional Minkowski space with 
one extra dimension.
Then we reduce the range of this coordinate at some finite value
and place a three-brane at the boundary.\footnote{The important issue of
the determination of the location of the brane from first principles 
will not be addressed in this paper. For the model in \cite{rasu1,rasu2} 
a suggestion, based on a modulus-field stabilization mechanism, 
was made \cite{wise2}; 
it can presumably be modified to cover our models as well.} 
Although the fifth coordinate is not necessarily compact, these
backgrounds are all effectively compactifications, since the
Kaluza--Klein spectrum contains 
a zero-mode corresponding to the four-dimensional gravity of our
world and there exists also a mass gap.
This mass gap can be chosen such that the corrections to Newton's law
lie within present-day experimental limits.
Therefore, an observer on the brane will effectively see 
a four-dimensional world.

We will concentrate on two backgrounds: 
one (to be called (A)) corresponds
to D3-branes distributed uniformly over a disc, with metric given by 
\be\label{5dbackg}
ds^2 = \frac{r^{4/3}}{R^2} (r^2 + r_0^2)^{1/3} \eta_{\m\n} dx^\m dx^\n +
\frac{R^2}{r^{2/3}} (r^2 + r_0^2)^{-2/3} dr^2\ ,\qq r\ge 0\ ,
\ee
and one (to be called (B)) corresponding to D3-branes distributed 
uniformly over a three-sphere, with metric given by
\be\label{5dbac2}
ds^2 = \frac{r^{4/3}}{R^2} (r^2 - r_0^2)^{1/3} \eta_{\m\n} dx^\m dx^\n +
\frac{R^2}{r^{2/3}} (r^2 - r_0^2)^{-2/3} dr^2\ , \qq r\ge r_0\ .
\ee
Note that the two metrics are related by the analytic continuation
$r_0 \to i r_0$ and $\eta_{\m\n}$ is the metric of four-dimensional 
Minkowski space-time. 

It is useful to present the metrics in their conformally flat form 
\be\label{5dconfflat}
ds^2 = e^{\displaystyle 2 \Omega(z)} ( \eta_{\m\n} dx^\m dx^\n + dz^2) \ .
\ee
For model (A) we find that the coordinate transformation 
\be\label{coord-change}
r = \frac{r_0}{ \sinh (z/R)} \ ,\qq z\ge 0\ ,
\ee
and a rescaling of the space-time coordinates as $x^\m \to {R\ov r_0}x^\m$
transforms \eqn{5dbackg} into the form \eqn{5dconfflat} with conformal factor
\be\label{warpdisk}
e^{\displaystyle \Omega} = {\cosh^{1/3} (z/R) \ov \sinh (z/R)} \ .
\ee
Similarly for model (B), the coordinate transformation 
\be\label{coord-chan1}
r = \frac{r_0}{ \sin(z /R)}\ ,\qq 0\le z\le {\pi R\ov 2}\ ,
\ee
and the same rescaling of the $x^\m$'s as before, 
transforms \eqn{5dbac2} into the form \eqn{5dconfflat} with conformal factor
\be\label{warpsphere}
e^{\displaystyle \Omega} = {\cos^{1/3} ( z/R)\ov \sin (z/R)}\ .
\ee
Near the boundary $r \to \infty$ ($z \to 0$) both spaces are asymptotic 
to $AdS_5$. Equivalently, we recover $AdS_5$ if we let the parameter
$R$ go to infinity. 

Our models are constructed by taking the above backgrounds
and cutting out the boundary region $0 \leq z \le z_1$ by placing a 
three-brane at $z=z_1$ (compare footnote 1). 
This restriction on the range of $z$ is necessary
to obtain dynamical gravity in the effective four-dimensional
theory. It is useful to consider the double cover of this space, 
which amounts to requiring reflection symmetry with respect to the three-brane.
A feature of these backgrounds is that the five-dimensional
bulk action also contains a scalar field with a potential (some
facts about five-dimensional supergravity relevant to our backgrounds are 
summarized in appendix A).
The five-dimensional action is the sum of a bulk term and
a boundary term: 
\ba
S & = & S_{\rm bulk} + S_{\rm brane} 
\nonumber \\
 & = & \int_{z_1}^{z_{\rm max}} dz \int d^4x \sqrt{-G} \left( \frac{1}{4} 
{\cal R} - 
\frac{1}{2} G^{MN} \partial_M \phi \partial_N \phi - P(\phi) \right) 
\nonumber \\
&&+ \int d^4x \sqrt{-\widetilde{G}}(\mathcal{L}_{\rm brane} - V)\ ,
\label{action}
\ea
with $z_{\rm max}=\infty$ in case (A) and $z_{\rm max}=\pi R/2$ in case (B).
The five-dimensional metric is denoted by $G_{MN}$, and ${\cal R}$ is 
the corresponding Ricci scalar. The potential $P$ of the scalar field 
$\phi$ is defined in appendix A, and $\widetilde{G}$
is the pullback of the bulk metric to the four-dimensional world
volume of the three-brane. 
The brane action
contains a term, $\mathcal{L}_{\rm brane}$, that corresponds to the
matter fields living on the brane and will be ignored in the following; 
the other term is $V$, which corresponds to the tension of the brane
and which will be fixed shortly by requiring consistency of the equations of 
motion. We will show that this also implies the vanishing of the 
effective cosmological term on the three-brane.

The equations of motion following from varying the metric and the scalar field
in the action \eqn{action} are
\ba
& & \frac{1}{4} R_{MN} - \frac{1}{2} \partial_M \phi \partial_N \phi -
\frac{1}{3} P G_{MN} - T_{MN} + \frac{1}{3}  T_L^L  G_{MN} = 0 \ ,
\nonumber\\ 
& & \partial_M \left( \sqrt{-G} G^{MN} \partial_N \phi \right)  - 
\sqrt{-G} \frac{\partial P}{\partial \phi} - \sqrt{-\widetilde{G}}
\frac{\partial V}{\partial \phi} \delta (z - z_1) = 0 \ ,
\label{eom}
\ea
where the energy--momentum tensor $T_{MN}$ of the source term coming 
from the three-brane is
\be 
T_{MN} = -\frac{V}{2}\frac{\sqrt{-\widetilde{G}}}{\sqrt{-G}} 
\widetilde{G}_{\m\n} \delta_M^\m \delta_N^\n \delta (z - z_1) \ .
\label{tnmas}
\ee
The solutions to these equations can be expressed in terms of the 
bulk solutions without the three-brane, 
by a simple replacement that preserves reflection
invariance $z - z_1 \to -(z -z_1)$ with respect to the brane
\be\label{rplc}
z \to \tilde{z} = z_1 + |z - z_1| \ .
\ee
This introduces terms proportional to $\delta (z-z_1)$ 
in the equations of motion, which are cancelled by the source term 
if we choose the vacuum energy of the three-brane to be 
\be\label{vacuum}
V =- 3 \Omega'(z) \exp(-\Omega) = \frac{2}{R} W(\phi) ~,
\ee
where $W$ is an auxiliary function of the scalar field $\phi$ and 
is defined in appendix A.

We will focus our attention to solutions of \eqn{eom} that preserve 
some supersymmetry, so that the corresponding backgrounds are stable, 
at least perturbatively.
The Killing spinor conditions for preserving sixteen supercharges, 
in the absence of the brane at $z=z_1$, were derived in \cite{FGPW} 
and are summarized in appendix A.
In our case it turns out that they are slightly
altered by the presence of the brane
\ba
\frac{d\Omega}{dz} & = & -\frac{2}{3 R} e^{\displaystyle \Omega} W \times 
\mathrm{sign}(z - z_1) \ ,
\nonumber\\
\frac{d\phi}{dz} & = & \frac{1}{R} e^{\displaystyle \Omega} 
\frac{\partial W}{\partial\phi} \times \mathrm{sign}(z - z_1) \ .
\label{dj3}
\ea
The solution is given by
the bulk solution after replacing $z \to \tilde{z}$ in 
\eqn{warpdisk} and \eqn{warpsphere}.

Now that we have found the backgrounds, we can proceed to calculate
several properties of the effective four-dimensional models.
In this paper we will concentrate on the gravity sector. 
This is partly because we do not know the exact form
of $\mathcal{L}_{\rm brane}$; of course, we could impose the standard
model Lagrangian, or try to derive it from first principles. In the
second case, keeping in mind the string theory origin of our
backgrounds, this would most probably amount to expanding the Born--Infeld 
action in the corresponding background. 
We leave this for future studies and ignore
for the moment contributions of the matter sector to the gravity equations. 
The quantities we want to study in the following are the four-dimensional
Planck constant and the four-dimensional cosmological constant, which
should vanish for a physical model.
In section 3 we will discuss the spectrum of linearized 
graviton fluctuations and in section 4 the corrections to Newton's law.

In order to calculate the four-dimensional Planck mass in terms of the
five-dimensional fundamental scale, we express the five-dimensional metric 
$G_{MN}$ in terms of a four-dimensional metric ${G}_{(4)\m\n}$, which replaces 
$\eta_{\m\n}$ in \eqn{5dconfflat} so that the various four-dimensional
geometrical data are non-vanishing. The effective four-dimensional 
Lagrangian thus derived from \eqn{action} is
\ba\label{4defflag} 
\mathcal{L}_{4d} & = & M^3 \int_{z=z_1}^{z_{\rm max}} \sqrt{- G_{(4)}} 
\left( \frac{1}{4} e^{\displaystyle 3 \Omega} R_{(4)} - 
3 e^{\displaystyle 3 \Omega} (\Omega')^2 - 
2 e^{\displaystyle 3 \Omega} \Omega'' -
\frac{1}{2} e^{\displaystyle 3 \Omega} (\phi')^2  \right. \nonumber \\
&&\left. - e^{\displaystyle 5 \Omega} P + \frac{3}{2} \Omega' 
e^{\displaystyle 3 \Omega} \delta (z-z_1) \right) \ ,
\ea
where $z_{\rm max}=\infty$ in case (A) and $z_{\rm max}=\pi R/2$ in case 
(B), and $M$ is the five-dimensional fundamental Planck constant.
The effective Planck constant in four dimensions, $M_{\rm Pl}$,
can be read off from the first term in \eqn{4defflag}. 
In case (A), we find 
\be\label{4dplanck}
M_{\rm Pl}^2 = M^3 \int_{z_1}^\infty dz~ 
e^{\displaystyle 3 \Omega(z)} = \frac{M^3 R}{2} \frac{1}{\sinh^2 (z_1/R)} \ ,
\ee
whereas in case (B)
\be\label{4dplanckb}
M_{\rm Pl}^2 = M^3 \int_{z_1}^{\pi R/2} dz~ 
e^{\displaystyle 3 \Omega(z)} = {M^3 R\ov 2} \cot^2(z_1/R) \ .
\ee
Because of the non-vanishing scalar potential $P(\phi)$, the five-dimensional 
model has an effective negative
cosmological constant that is $z$-dependent. Hence, it is not
obvious that this will not induce an
undesirable cosmological constant in four dimensions (the latter, if 
non-vanishing, can be consistently set to zero 
for the purposes of this paper).
However, this is prevented even by mathematical consistency, since it 
would imply that the metric 
on the three-brane would not be a Minkowski but a curved one.
Indeed it may be checked that although all terms in \eqn{4defflag}, except 
the first one, contribute to the cosmological constant,
the final result is zero.

\subsection{The parameter space}

Our models have three parameters: the fundamental five-dimensional mass scale
$M$, the position of the brane $z_1$, and the length parameter $R$.
These have to be chosen in such a way that supergravity is a good 
approximation as an effective theory. A straightforward computation for the 
scalar curvature for the metric \eqn{5dconfflat} gives
\be
\mathcal{R} = -4 e^{2 \Om} \left(2 \Om^{\prime\prime} + 3 \Om^{\prime 2}
\right)\ .
\label{dh11}
\ee
Using the explicit expressions for the conformal factors 
\eqn{warpdisk} or \eqn{warpsphere}, it turns out that for ratios $z/R$ 
not too close to $z_{\rm max}$,
the scalar curvature 
becomes $\mathcal{R} \sim -{\rm const.}/R^2$, where the proportionality 
constant is of order $1$. A similar statement holds 
for the other curvature invariants.
Hence, the condition $|\mathcal{R}|\ll M^2 $ for supergravity to be valid 
reduces to 
\be 
R M \gg 1 \ .
\label{h34}
\ee 
Keeping this in mind, we investigate the different physical pictures 
that are obtained from various choices of the 
location of the three-brane $z_1$, as well as of the 
fundamental five-dimensional scale $M$ and the length parameter $R$.

\no
{\bf Case I}: The three-brane is located at the point where
\be
e^{\Omega(z_1)} = 1\ ,
\label{hj3}
\ee
which implies, using the background \eqn{5dconfflat}, that $z_1 = 0.984 R$ 
when the conformal factor is given by \eqn{warpdisk} and $z_1 = 0.972 R$ 
when the conformal factor is given by \eqn{warpsphere}.
We take the radius of curvature of our background to be of the order
of $1$ mm and find that
\be
R \sim 1\ \mathrm{mm} \sim 10^{14}~ \mathrm{GeV}^{-1} \ ,\qq
M \sim 10^8~ \mathrm{GeV}~,
\label{hdf}
\ee
where the five-dimensional fundamental mass scale was found 
using \eqn{4dplanck}. Notice that this value is intermediate between the 
Planck and electroweak energy scales. 
The picture that emerges is similar to the case of compactification
on a large torus of radius $\sim 1$ mm. Consistent with that is the fact that
both \eqn{4dplanck} and \eqn{4dplanckb} can be approximated by
\be
M^2_\mathrm{Pl} \sim M^3 R\ .
\label{hb5}
\ee
Similarly to \cite{AHDD} (for the case of compactification on a large torus), 
unification of the fundamental scale $M$ and the electroweak scale 
$M_{\rm E}$ cannot be achieved 
unless $R \sim 10^{13}$ m, i.e. $R$ has astronomical size. 
Choosing $M$ at an intermediate scale, as in \eqn{hdf}, 
we avoid this problem and we are consistent with present-day data.

\no
{\bf Case II}: The three-brane is located at the point where
\be
e^{\Omega(z_1)} \sim \frac{R}{z_1} \sim 10^{11}\ ,
\label{hj4}
\ee
and $M$ and $1/z_1$ are both taken to be of the order of the electroweak scale
$\sim 1\ \mathrm{TeV}$. Since the argument of the conformal factor is small
the four-dimensional Planck constant can be approximated as
\be
M^2_\mathrm{Pl} \sim \frac{M^3}{z_1^2} R^3\sim M^5 R^3 \ ,
\label{fj3}
\ee
which is similar to compactifications with three extra dimensions
of large size $R \sim 10^{-5}~ \mathrm{mm} \sim 10^{8}~ (\mathrm{GeV})^{-1}$.
In this model the wave functions of the massive gravitons 
are extremely small, at $z=z_1$. For this reason, as 
we shall see
in section 4, the Yukawa-type corrections to Newton's law turn out to be
negligible.

We note that for the choice of parameters we have made above, the condition 
\eqn{h34} is clearly satisfied.
A feature that the two geometries share is that the curvature 
blows up close to the brane distribution (at $z\to \infty$ and $z\to \pi/2 R$
for model (A) and (B) respectively) and corrections due to 
higher-derivative terms in the action become important.
However, the region of large curvature is very small 
and the qualitative picture will not be altered at all. 
This point was discussed in \cite{FGPW,BS} 
from the point of view of the ten-dimensional D3-brane solution.

\section{The graviton: massless and massive modes}

In this section we study small fluctuations $h_{\m\n}$  
of the four-dimensional Minkowski metric on the brane 
and determine the graviton spectrum. 
In general, this will depend on the
details of the five-dimensional backgrounds where the 
four-dimensional Minkowski metric is embedded.

We parametrize the fluctuations in the following manner
\be\label{5dfluct}
ds^2 = e^{\displaystyle 2 \Omega(\tilde{z})} 
\left( (\eta_{\m\n} + h_{\m\n}) dx^\m dx^\n + dz^2 \right) \ ,
\ee
as it is consistent to set to zero the components of the graviton
fluctuations $h_{5\m}$
and $h_{55}$ as well as the fluctuations of the scalar field $\phi$.
Then we insert \eqn{5dfluct} into \eqn{eom} and linearize in $h_{\m\n}$. The
calculation is facilitated by the fact that the metric is
conformally flat. We may also utilize the reparametrization invariance 
by choosing the gauge $h^\m_\m = \partial^\m h_{\m\n} = 0$. In this
gauge we find the following equation:
\be\label{flucteq} 
\square_x h_{\m\n} + \partial^2_z
h_{\m\n} + 3 \frac{d\Omega(\tilde{z})}{dz}  \partial_z h_{\m\n} = 0\ ,
\ee
which is just the Laplace equation of a massless scalar in the five-dimensional
background \eqn{5dfluct}. The source term due to the three-brane cancels out
completely, but the presence of the brane at $z=z_1$ demands that 
appropriate boundary conditions are chosen, as we shall see. 
We consider fluctuations that are plane waves in the four-dimensional
Minkowski space-time
\be
h_{\m\n}(x,z)= \exp(i k \cdot x) h_{\m\n}(z)\ .
\label{graa}
\ee
Then 
we find the following simple differential equation (we can drop the indices
from $h_{\m\n}$ since the differential equation is the same for all the
components of the graviton):
\be\label{waveeq}
h^{\prime\prime} + 3 \frac{d\Omega(\tilde{z})}{dz}  h^\prime + 
M^2 h = 0\ ,
\ee
where $M^2 = -k\cdot k$ is the mass square of the corresponding
graviton fluctuation mode.
At the location of the three-brane, the wave function $h$ and
its first derivative are smooth functions. This implies, 
because of the reflection symmetry with respect to the three-brane, that the
first derivative of $h$ has to vanish at the location of the three-brane.
This condition also guarantees the hermiticity of the Laplacian in the curved
background \eqn{5dfluct}.
It is obvious that there exists a massless mode
\be
h_0 = {\rm const.}
\label{gg1}
\ee
that solves \eqn{5dfluct} and also has vanishing
first derivative at the position of the brane at $z=z_1$. 
Moreover, it is normalizable in the interval $z_1 \leq z < z_{\rm max}$, 
with measure $dz e^{3 \Om}$.
In the rest of this section we determine the spectrum of the massive graviton 
fluctuations for our two models.

\subsection{Model A}

We first determine the
mass spectrum of graviton fluctuations for non-zero $M$ 
for the model corresponding to a uniform distribution of D3-branes on a disc.
In order to solve the differential equation we found it useful to
change variable to 
\be
x= {U^2\ov U^2+r_0^2} = {1\ov \cosh^2(z/R)}\ ,\qq 0\leq x\leq x_1\equiv 
{1\ov \cosh^2(z_1/R)}< 1\ .
\label{jds}
\ee
Then the function $hx^{(1-q)/2}$ obeys 
a hypergeometric equation. Hence, we can easily 
write down the general solution for the graviton fluctuations as 
\ba
\label{n2solution}
h_q & =& {\cal N}_q \Big( e^{i \th_q}\ x^{(q-1)/2} F_q(x) 
\ +\  e^{-i \th_q}\ x^{-(q+1)/2} F_{-q}(x) \Big)\ .
\label{duw}
\ea
The constant $q$ and the function $F_q(x)$ are related 
to the mass $M$ and a particular hypergeometric function as
\ba 
F_q(x) & \equiv& F\Big({q-1\ov 2},{q-1\ov 2},1+q; x\Big)\ ,
\nonumber \\
q & \equiv & \sqrt{1-R^2 M^2} \ ,
\label{gqe}
\ea
where ${\cal N}_q$ and $\th_q$ are real constants.
The above solution takes the same form as that given in \cite{BS}, after using 
certain transformation properties of hypergeometric functions.
The coefficients and the allowed values of $M$ are now determined 
by requiring normalizability and the appropriate boundary
conditions.
Since $x$ does not take values in the entire unit interval, the hermiticity 
condition for the scalar Laplace operator requires that 
\be
\left. {\del_x h}\right|_{x=x_1} = 0 \ .
\label{der1}
\ee
This determines the phase $\th_q$ as
\be
e^{2 i \th_q} = {(q+1) F_{-q}(x) - 2 x \del_x F_{-q}(x)\ov 
(q-1) F_{q}(x) + 2 x \del_x F_{q}(x)}\Bigg |_{x=x_1}\ .
\label{psaa}
\ee
Note that $\th_q$ depends not only on the value of $q$, but also on the 
particular point $x_1$. 
As was shown in \cite{BS} (see also \cite{FGPW}) the parameter
$q$ in \eqn{gqe} is purely imaginary and therefore
there is a mass gap in the spectrum 
\be
M_{\rm gap}= {1\ov R}\ ,
\label{gapp}
\ee
and, above it, there is a continuum with $M \geq 1/R$.
This follows from
requiring orthonormalizability in the Dirac sense (with the use of a 
$\d$-function). 
We also note that the existence of a mass gap is most transparent 
if we transform the equation for the graviton fluctuations $h(z)$ into 
a Schr\"odinger equation by means 
of the transformation $h(z) = e^{-3 \Omega/2} \Psi(z)$
\be
-{d^2\Psi\ov dz^2} + V(z) \Psi = M^2 \Psi\ , 
\label{sch1}
\ee
with potential 
\ba
V(z)  = {3\ov 2} \Omega^{\prime\prime}(z) +{9\ov 4} \Omega^{\prime 2}(z) 
+ 3 \Omega^\prime  (z_1) \d(z-z_1)\ .
\label{jdgg}
\ea
In our case, using \eqn{warpdisk}, we obtain 
\be
V(z) = {1\ov R^2} \left(1+ {1\ov 4 \cosh^2(z/R) }
+{15\ov 4 \sinh^2(z/R)} \right) 
-{2\ov R}  {2+ \cosh(2 z_1/R)\ov \sinh(2 z_1/R)} \d(z-z_1)\ ,
\label{jd}
\ee
which is a particular member of the class of potentials known in the 
literature as P\"oschl--Teller potentials of type II.
Clearly, since $V(z\to \infty)=1/R^2$, there is a minimum value for $M^2$ 
given by \eqn{gapp}. Note also that the value of the mass gap does not
depend on the particular location of the brane at $z_1$. The reason is 
that the mass gap corresponds to the asymptotic value of the potential
for large $z$. The zero-mode wave function is 
\be 
\Psi_0(z) = {\cosh^{1/2}(z/R)\ov \sinh^{3/2}(z/R)}\ ,
\label{jasg}
\ee
and is clearly normalizable in the interval $z_1\leq z < \infty$. 
Of course it corresponds to \eqn{gg1} after we multiply with the 
factor $e^{3\Om/2}$.

\subsection{Model B}

Let us now turn to the case of the model corresponding to D3-branes 
uniformly distributed over a three-sphere. 
In order to solve the differential equation \eqn{waveeq} with \eqn{warpsphere}
we found it useful to change variable to
\be
x= 1-{r_0^2\ov U^2} = \cos^2(z/R)\ ,
\qq 0\leq x\leq x_1\equiv \cos^2(z_1/R)< 1\ .
\label{jds1}
\ee
Then the function $(1-x)^{-2}h$ obeys 
a hypergeometric equation. The general solution for the graviton fluctuations 
that is regular at $x=0$ is given by
\be
h_q = \tilde {\cal N}_q (1-x)^2 F(q+2,-q+1,1;x)\ ,
\label{jd21}
\ee
where $\tilde {\cal N}_q$ is a normalization constant and 
the real number $q$ parametrizes the mass $M$ as $M= {2\ov R} \sqrt{q(q+1)}$.
The hermiticity condition for the scalar Laplace operator is equivalent to
the condition \eqn{der1} for the derivative of the graviton wave function.
This determines the massive spectrum as
\be
M_q= {2\ov R} \sqrt{q(q+1)}\ ,%\qq n=1,2,\dots \ .
\label{hjq}
\ee
where $q$ belongs to a discrete set of real numbers that can be taken
to be positive with no loss of generality.
Hence, as in the case of the previous model, there exists a mass gap, 
corresponding to the lowest eigenvalue 
for $q$, separating the massless graviton mode from the massive 
spectrum.\footnote{For
$z_1 = 0.972 R$ (equivalently $x_1 = 0.318$) we have 
$e^{\Omega(z_1)} = 1$ and the first few eigenvalues are  $q = 2.74, 5.24, 
8.00, 10.6\dots $ As the brane position $z_1$ moves closer to $z=0$ 
(equivalently $x=1$) the 
corresponding eigenvalues move towards the set of positive integers
(see \eqn{qi21} and \eqn{has1} below).
Hence, the mass gap corresponding to the lowest eigenvalue for $q$
is insensitive, for all practical purposes, to the position of the 
brane.}
As before we may cast our eigenvalue problem into an equivalent 
Schr\"odinger problem. The corresponding potential is 
given by \eqn{jdgg} after we use \eqn{warpsphere}
\be
V(z) = {1\ov R^2} \left(-1- {1\ov 4 \cos^2(z/R) }
+{15\ov 4 \sin^2(z/R)} \right) 
-{2\ov R} {2+ \cos(2 z_1/R)\ov \sin(2 z_1/R)} \d(z-z_1)\ .
\label{j1d}
\ee
Similarly to \eqn{jd}, it is a particular member of the class of potentials 
known in the 
literature as P\"oschl--Teller potentials of type I. 
The zero-mode wave function for the potential \eqn{j1d} is given by
\be 
\Psi_0(z) = {\cos^{1/2}(z/R)\ov \sin^{3/2}(z/R)}\ ,
\label{jasg1}
\ee
which is normalizable in the interval $z_1\leq z < \pi R/2$. 
Of course it corresponds to \eqn{gg1} after we multiply it by the  
factor $e^{3 \Omega/2}$.
The Schr\"odinger differential equations for the potentials
\eqn{jd} and \eqn{j1d} are related by the analytic continuation 
$z\to i z$, as expected from
a similar relation between the corresponding supergravity backgrounds.

In general the wave functions \eqn{jd21} 
are not orthogonal for different values of 
$q$ and in addition they overlap with the massless constant mode $h_0$.  
In principle it is straightforward to construct an orthonormal basis, 
although in practice one eventually resorts to numerical methods.
There are, however, two particular choices for the position
of the brane that lead to a great mathematical simplification
and make the physical picture more transparent.
In the first case, consider $x_1\simeq 1$. This corresponds to $z_1\ll R$
as in case II discussed in subsection 2.1.
In this case, \eqn{jd21} (with $q$ a positive integer) 
is related to Jacobi polynomials. The complete set of mutually 
orthogonal massive graviton 
wave functions that are normalized to 1 is practically the same as the one
found in \cite{BS}, where $x_1=1$ exactly. They read 
\be
h_n = \sqrt{2(2 n+1)\ov R} (1-x)^2 P_{n-1}^{(2,0)}(2 x-1)\ ,
\qq 0\leq x \leq 1\ , \qq n=1,2,\dots \ ,
\label{qi21}
\ee
where $P_{2n-1}^{(2,0)}$ are the appropriate Jacobi polynomials. The mass 
spectrum is, to an extremely good approximation, given by 
\be 
M_n  = {2\ov R} \sqrt{n(n+1)}\ ,%\ e^{-A(z_1)}\ , 
\qq n=1,2,\dots \ .
\label{has1}
\ee

Another case where Jacobi polynomials arise is for $x_1=1/2$
(which corresponds to $z_1 = \pi R/4 \sim 0.785 R$ and $e^{\Omega(z_1)} =
2^{1/3} \sim 1.26$). 
Note that for this choice of the brane location we are effectively discussing
a parameter choice similar to that made in case I in subsection 2.1.
Here, the set of real numbers $q$ in \eqn{jd21} coincides with the set 
of even integers. It turns out that the 
complete set of mutually orthogonal massive graviton 
wave functions that are normalized to 1 is 
\be
h_n = 2 \sqrt{4 n+1\ov R } (1-x)^2 P_{2n-1}^{(2,0)}(2 x-1)\ ,
\qq 0\leq x \leq \ha\ , \qq n=1,2,\dots \ ,
\label{qi1}
\ee
whereas the mass spectrum is given by 
\be 
M_n  = {4\ov R} \sqrt{n(n+1/2)}\ ,
\qq n=1,2,\dots \ .
\label{has}
\ee
The set of eigenfunctions \eqn{qi1} is by itself complete and 
all of its members are orthogonal 
(with respect to the usual inner product) to 
the constant massless mode. The same of course is true for \eqn{qi21}.

\section{The Newton law}

The fact that we have modelled our flat four-dimensional space-time as
a brane embedded in a five-dimensional curved space-time, has certain
consequences for the 
Newton law that governs the gravitational
attraction of point particles in our four-dimensional world. 
Present-day experimental data do not exclude
corrections to the $1/r^2$ attractive force for distances smaller than 
or equal to $1$ mm. Since our models are five-dimensional at the 
fundamental level, we should, at distances much smaller than $R$, have 
the Newton law in four spatial dimensions 
\be
V= - {G_5 M\ov r_4^2}\ ,
\label{nee4}
\ee
where $r_4$ denotes the corresponding radial distance from a point mass $M$
located at the origin, and $G_5$ is the Newton constant in $4+1$ dimensions.
On the other hand, for distances much larger than $R$, we should just obtain
the usual Newton law in three spatial dimensions
\be
V= - {G_4 M\ov r}\ ,
\label{nee2}
\ee
where $r$ is the usual radial distance in three spatial dimensions
and $G_4$ is the Newton constant in $3+1$ dimensions.
The crossover behaviour between \eqn{nee4} and \eqn{nee2} depends on the 
details of the particular model used. In particular, in view of 
possible experimental verifications, we are interested in computing exponential
corrections to the leading-order behaviour of the Newton potential \eqn{nee2}.
At this point we emphasize again the importance of the existence of a mass gap 
in our models, as this will govern the behaviour of the leading Yukawa-type
correction to \eqn{nee2}.\footnote{ The model in \cite{rasu2}, 
based on the $AdS_5$,
has a continuous spectrum with no mass gap, and the
corrections to Newton's law are power-like. They can be made extremely small 
when the curvature of the five-dimensional space-time is of Planck size
\cite{rasu2}.}

At the linearized level the 
gravitational potential in 4+1 dimensions with curved metric given by 
\eqn{5dconfflat} obeys the following equation 
in four spatial dimensions
\be\label{fl1cteq}
\nabla_3^2 V + \partial^2_z V + 3 \frac{d\Omega(\tilde{z})}{dz}  \partial_z V 
= 4\pi^2 M G_5 {\d^{(3)}({\bf x})
\d(z-z_1)\ov e^{3 \Omega}}\ .
\ee
This is nothing but \eqn{flucteq}, after we include 
a source term, due to a point particle of mass $M$ located at the brane and 
at the origin of our three-dimensional spatial world. 
The normalization in the right-hand side of \eqn{fl1cteq} is chosen such that 
for small distances the potential is $V= - {G_5 M\ov r^2+ (z-z_1)^2}$, 
in accordance with \eqn{nee4}.
We have also dropped the time dependence since we are seeking static solutions.
We may expand $V$ in terms of the complete basis of
eigenfunctions $\{h_Q\}$ of the operator $\del_z^2 + 3 d\Omega(\tilde{z})/dz 
~\del_z$
as
\be 
V= \sum_{Q} V_Q(r) h_Q(z)\ ,
\label{jan}
\ee
where the sum over $Q$ comprises the massless 
as well as the massive modes. 
The general solution can be written formally as
\be 
V(r) = - {\pi G_5 M\ov r} \sum_{Q,P} e^{-M_Q r} K\inv_{QP} h_Q(z_1)
h^*_P(z_1)\ ,
\label{sol1}
\ee
with 
\be
K_{QP} = \int_{z_1}^{z_{\rm max}} dz 
e^{\displaystyle 3 \Omega(\tilde{z})} h_Q^*(z) h_P(z)\ ,
\label{dj1}
\ee
with $z_{\rm max} = \infty$ and $\pi R/2$ for the case of our models (A) and 
(B). 
We have omitted the internal space dependence 
since all point particles are at $z_1$ with respect to 
our four-dimensional space-time. 
In general, we may assume that we have an orthonormal system of eigenfunctions
$\{h_Q\}$. Then \eqn{sol1} becomes
\be 
V(r) = - {G_4 M\ov r} \left(1+ \sum_{q} d_q(z_1) e^{-M_q r}\right)\ ,\qq
d_q(z_1)\equiv |h_q(z_1)|^2 \int_{z_1}^{z_{\rm max}} dz e^{3 \Om}\ ,
\label{rql1}
\ee
where the sum is now over the massive modes only and the four- and 
five-dimensional Newton constants are related as
\be
G_4 = {\pi G_5\ov \int_{z_1}^{z_{\rm max}}dz e^{3\Om}}\ .
\label{qhw}
\ee
The factor $d_q(z_1)$ in \eqn{rql1}
weights the contributions of the various massive
Kaluza--Klein states.\footnote{In fact \eqn{rql1} is very similar and 
extends the corresponding formulae proved in
\cite{kesf} for compactification on general compact manifolds (for
the case of a torus compactification, see also \cite{FL}).
One apparent difference 
is that, in the corresponding formulae in \cite{kesf}, the 
degeneracy of the irreducible representations 
of the symmetry group of the compact space appears instead of 
$d_q(z_1)$. 
However, as shown in \cite{kesf}, 
this degeneracy can be also written in terms of eigenfunctions of the 
scalar Laplacian on the compact internal space.}
In practice it is difficult to construct an orthonormal basis, 
as can be seen from our examples \eqn{duw} and \eqn{jd21}.
However, we may easily deduce that the general form of the Newton 
potential, with the leading correction included, will be 
\be
V(r)\simeq -{G_4 M\ov  r} \left(1+ \a(r) e^{-r/\l}\right)\ ,
\label{ehj}
\ee
where the range $\l$ of the Yukawa correction is related to the
mass gap. The strength of the force $\a(r)$ could itself be a 
function of the distance $r$, as indicated,  provided that the spectrum above
the mass gap is continuous. However, if the spectrum is discrete the strength 
is just a constant.
For the case of our examples we may show that 
\be
{\rm model\ (A):}\qq\qq \l= R \ ,\qq \a(r) ={\rm const.}
\left(R\ov r\right)^2\ ,
\label{ejh1}
\ee
and 
\be
{\rm model\ (B):}\qq\qq \l\sim {R} \ ,
\qq \a(r)= {\rm const.}\ , \qq ~~~
\label{ejh2}
\ee
where both constants appearing in the previous expressions depend on 
the position of the brane. For the choice of parameters in case I in subsection
2.1 both of these constants are of order 1, resulting in a measurable
contribution to the strength of the Yukawa correction.
Unlike this case, for the choice of parameters in case II in subsection
2.1 both of these constants are extremely small due to the wave function 
suppression. Hence, the corresponding Yukawa correction is negligible.
We may demonstrate these different behaviours using 
the two orthonormal sets of wave functions \eqn{qi1} and \eqn{qi21} 
as they correspond to two extreme choices of the brane location.

For the special case of the orthonormal wave functions \eqn{qi1} 
with masses given by \eqn{has}, the explicit computation can be carried out 
without much effort using properties of the Jacobi polynomials. We find that 
\ba
V(r) & = & -{G_4 M\ov  r} \left( 1 + \sum_{n=1}^\infty 
A_n e^{-{4 r\ov R} \sqrt{n(n+1/2)}} \right) \ , 
\nonumber\\
A_n & = & {4 n+1\ov 2^{2n+1} n^2} \left({(2n-1)!!\ov (n-1)!}\right)^2 \ .
\label{jwdb}
\ea
In the limit of large $r$ we may keep only the first term in the infinite
sum above to a very good approximation 
\be
V(r)  \simeq  -{G_4 M\ov r}\left(1+ {5\ov 8} e^{-{2 \sqrt{3}\ov R} r}\right)\ ,
\qq {\rm for}\quad r\gg R\ .
\label{lii2}
\ee
In the opposite limit of small $r$ we may  
approximate $A_n$ in \eqn{jwdb} by its value at infinity,
i.e. $A_n\simeq A_{\infty}=2/\pi$. Using also the relation $2 \pi G_5= R G_4$
(following from \eqn{qhw}) we indeed obtain the potential corresponding
to Newton's law in four spatial dimensions: 
\be
V(r)\simeq -{G_5 M\ov r^2} \ , \qq {\rm for}\quad r\ll R\ ,
\label{je2}
\ee
as expected. The advantage of \eqn{jwdb} is that the crossover behaviour from 
\eqn{je2} to \eqn{lii2} is expressed in a simple and precise way.

For the special case of the orthonormal wave functions \eqn{qi21} 
with masses given by \eqn{has1} we find that 
\ba
V(r) & = & -{G_4 M\ov  r} \left( 1 + \sum_{n=1}^\infty 
B_n e^{-{2 r\ov R} \sqrt{n(n+1)}} \right) \ , 
\nonumber\\
B_n & = & {(1-x_1)^3\ov 4} (2 n+1) n^2 (n+1)^2 \ + \ {\mathcal O}(1-x_1)^4\ .
\label{jwd}
\ea
Hence, we see that using \eqn{hj4} and \eqn{jds1} all corrections 
to Newton's $1/r$ potential are ${\mathcal O}(10^{-66})$ and therefore
negligible. As we have mentioned, this is due to the wave-function suppression
near $x=1$.

\section{Discussion}

In this paper we have considered models of our four-dimensional world 
in which a three-brane is embedded in five-dimensional backgrounds, 
which are solutions of gauged supergravity with a boundary term.
These backgrounds arise as consistent truncations of solutions of 
type-IIB string theory, which correspond to continuous distributions of
D3-branes on a disc or on a three-sphere.
From the gauged supergravity point of view, the 
five-dimensional metrics are warped products of Minkowski space
in four dimensions with an extra dimension. 
The interesting feature
of our backgrounds is that the spectrum of linearized  graviton 
fluctuations either has a genuine mass gap with a continuum above it or
it is discrete.
Furthermore, if the range of the fifth coordinate is 
reduced by placing a three-brane, a normalizable zero-mode appears, 
corresponding to the graviton in our world 
while the mass gap is preserved.
Therefore, these models naturally lead to physics that appears
four-dimensional, as long as energies are smaller than the mass gap,
and an effective four-dimensional theory can be defined.

Furthermore, 
we studied in detail the corrections of Newton's law induced by the 
massive excitations that are Yukawa-like owing to the mass 
gap. We examined the space of parameters, under the condition
that supergravity is a good approximation, for physically 
interesting scenarios. 
In one of them (case I in section 2.1) 
the fundamental scale is between the Planck and the
electroweak energy scales and the radius of curvature of the space 
is of millimetre size.
In this case the corrections are close to the experimental bounds and might
be observed in future experiments; but 
still there exists a hierarchy, which, however,  
is  smaller than the usual $10^{16}$. 
In this range of parameters our models resemble those in \cite{AHDD} with 
a single compactified large extra dimension. 
The other scenario (case II in section 2.1) has the attractive feature that
the fundamental scale is actually of the same order as the electroweak scale 
(unification) 
and it shares several features of compactification with three large extra
dimensions, although our models have only one.
The corrections to Newton's law, however, turn out to be negligible, 
because of a wave function suppression at the location of the three-brane.
It would be very interesting to find models similar to the ones
studied in this paper 
where the parameters can be chosen such that the corrections are 
closer to the experimental bounds whereas the fundamental scale can be
of the order of TeV, i.e. we have unification.

\bigskip

\centerline{\bf Note added}

While we were finishing up our paper, \cite{HDDK} appeared, where the idea
that the model of \cite{rasu1,rasu2} may correspond, 
with appropriate choice of parameters, to a direct product 
compactification with one large extra dimension, was also pointed out.

%%%%%%%%%%%%%%%%%%%%%%%%%%%%%%%%%%%%%%%%%%%%%%%%%%%%%%%%%%%%%%%%%%%%%%
%%%%%%%%%%%%%%%%%%%%%%%%%%%%%%%%%%%%%%%%%%%%%%%%%%%%%%%%%%%%%%%%%%%%%%

\appendix
\section{D3-branes distributions and gauged supergravity}

\setcounter{equation}{0}
\renewcommand{\theequation}{\thesection.\arabic{equation}}

The general ten-dimensional metric of an arbitrary
distribution of parallel three-branes in type IIB string
theory takes the following form
\be
ds^2 = H^{-1/2} \eta_{\m\n} dx^\m dx^\n + H^{1/2}(dr^2+r^2 
d\Om_5^2)\ ,\qq \m,\n=0,1,2,3\ ,
\label{background1}
\ee
where $d\Om_5^2 $ denotes the line element for the five-dimensional sphere
and the function $H$ is a harmonic function on the six-dimensional space 
transverse to the brane. 
One of the simplest examples is a stack of $N$ coinciding 
branes in the
near horizon limit: $H = R^4/r^4$ with $R^4 = 4 \pi g_s N l_s^4$.
In this case the background is $AdS_5 \times S^5$, which plays a central 
role in the AdS/CFT correspondence \cite{malda}. Upon reduction on the
five-sphere, this yields  ${\mathcal N}=8$ gauged supergravity in 
five dimensions on $AdS_5$. 
The authors of \cite{rasu1,rasu2} utilized 
variants of this background by adding one or two three-branes,
which cut off the fifth coordinate. 
In the full $AdS_5$ background, the gravity decouples from the 
four-dimensional
world but, once the fifth coordinate does not run over the
full range, gravity becomes dynamical. 
The $AdS_5$ solution in the bulk preserves thirty-two supersymmetries 
and is a stable background.

The metrics \eqn{background1} with harmonic $H$ corresponding to
more general not-coinciding three-branes provide a large class of 
interesting backgrounds for compactification, which preserve
sixteen supercharges. 
In the following we will concentrate on two backgrounds: (A) D3-branes
distributed uniformly over a disc and
(B) D3-branes distributed uniformly over a three-sphere. 
For (A) the metric can be written as \cite{BS} 
\ba
&& ds^2  = {r (r^2+r_0^2 \cos^2\th)^{1/2}\ov R^2} \eta_{\m\n} dx^\m dx^\n\
+\ {R^2\ov  r (r^2+r_0^2 \cos^2\th)^{1/2}}
\nonumber\\
&& 
\times \left( (r^2+r_0^2\cos^2\th) 
\Big({dr^2\ov r^2+r_0^2}+ d\th^2 \Big)
+ (r^2+r_0^2) \sin^2\th d\phi^2 
+r^2 \cos^2\th d\Omega_3^2\right)\ ,
\label{ruu1}
\ea 
where $d\Om_3^2 $ is the line element of the three-sphere, 
$r_0$ is the radius of the disc and $r \ge 0$. 
The metric for the case (B) is obtained by taking $r_0 \to - i r_0$
and restricting the range of the radial coordinate $r \ge r_0$.

These two ten-dimensional backgrounds can be consistently truncated to 
five-dimensional gauged supergravity.
More technical details can be found in \cite{FGPW}. 
The situation can be summarized as follows. 
The ${\mathcal N} =8$ gauged supergravity in five dimensions contains 
forty-two scalars, which have a non-trivial potential.
There is a stationary point where all scalars are zero 
(except for the complex coupling, which is a flat direction
of the potential), which
corresponds to the $AdS_5 \times S^5$ background in ten dimensions. 
For general D3-brane configurations the metric on $S^5$ is deformed, the
ten-dimensional space becomes a warped product space and in the
five-dimensional perspective 
(some of) the forty-two scalars develop non-trivial profiles. 
It is widely believed that
every five-dimensional solution can be lifted unambiguously
to ten dimensions, but a complete proof is still missing. 
There exist privileged flows such that 
the scalars lie in a one-dimensional submanifold of the forty-two
scalars, which we will denote by a scalar $\phi$.
It was shown in \cite{FGPW} that these solutions correspond to certain
D3-brane distributions in ten dimensions -- including our two examples.

The relevant part of the bosonic action of the five-dimensional gauged 
supergravity, 
in $-++++$ signature, is
\be\label{5dsugra}
{\cal L} = \sqrt{-G} \left( \frac{1}{4} {\mathcal R} - \frac{1}{2} G^{MN} 
\partial_M \phi \partial_N \phi - P(\phi) \right) ~,
\ee
where $\phi$ is the scalar field with potential $P$, 
which can be expressed in terms of an auxiliary 
function $W$ 
\be\label{5dpotential}
P = \frac{1}{2 R^2}  \left( \frac{\partial W}{\partial \phi} \right)^2
- \frac{4}{3 R^2} W^2 ~.
\ee
The five-dimensional metric has the form
\be\label{5dmetric}
ds^2 = e^{\displaystyle 2 \Omega(z)}( \eta_{\m\n} dx^\m dx^\n + dz^2) ~.
\ee
If the following two conditions: 
\be\label{5dshift}
\frac{d\phi}{dz} = \frac{1}{R} e^{\displaystyle \Omega} 
\frac{\partial W}{\partial\phi}~,~~~~~
\frac{d\Omega}{dz} = -\frac{2}{3 R} e^{\displaystyle \Omega} W ~,
\ee
are obeyed, then the solution preserves sixteen supercharges.
Solutions of \eqn{5dshift} automatically fulfil the field
equations of \eqn{5dsugra}.

We summarize here the results for our two examples \cite{BS,FGPW}.
For D3-branes smeared over a disc (case (A)) we have
\ba
W & = & e^{2 \phi/\sqrt{6}} + \frac{1}{2} e^{-4 \phi/\sqrt{6}} ~,
\nonumber \\
\Omega & = & -\frac{1}{2} \ln \left| e^{-2 \phi/\sqrt{6}} - 
e^{4 \phi/\sqrt{6}} \right| + \ln (r_0/R) ~, 
\label{wa-disk} \\
P & = & -\frac{1}{R^2} \left( e^{4 \phi/\sqrt{6}} + 2 e^{-2 \phi/\sqrt{6}}
\right) ~,
\nonumber
\ea
where $\phi$ is a scalar field depending on $z$.
The functions $W$, $\Omega$ and the potential $P$ for the 
case (B) can be obtained by replacing $\phi \to -\phi$.
Equation \eqn{5dshift} can
be solved for $\phi$. In the case (A) the solution is 
\be\label{solmua}
\phi = \sqrt{2 \ov 3} \ln \cosh \left({z r_0\ov R^2}\right) ~,
\ee
while in case (B) it is 
\be\label{solmub}
\phi = \sqrt{2 \ov 3} \ln \cos \left( \frac{z r_0}{R^2} \right) ~.
\ee

%%%%%%%%%%%%%%%%%%%%%%%%%%%%%%%%%%%%%%%%%%%%%%%%%%%%%%%%%%%%%%%%%%%%

%\newpage


\begin{thebibliography}{3}

\bibitem{GW}
V.A.~Rubakov and M.E.~Shaposhnikov,
Phys. Lett. {\bf 125B} (1983) 136;
%%CITATION = PHLTA,125B,136;%%
G.W.~Gibbons and D.L.~Wiltshire, Nucl. Phys. {\bf B287} (1987) 717 and
references therein.

\bibitem{rasu1}
L.~Randall and R.~Sundrum,
{\it A large mass hierarchy from a small extra dimension},
{\tt hep-ph/9905221}.
%%CITATION = HEP-PH 9905221;%%

\bibitem{rasu2}
L.~Randall and R.~Sundrum,
{\it An alternative to compactification},
{\tt hep-th/9906064}.
%%CITATION = HEP-TH 9906064;%%

\bibitem{AHDD} N.~Arkani-Hamed, S.~Dimopoulos and  G.~Dvali,  
Phys. Lett. {\bf B429} (1998) 263,  {\tt hep-ph/9803315};
I. Antoniadis, N. Arkani-Hamed, S. Dimopoulos and  G.  
Dvali,  Phys. Lett. {\bf B436} (1998) 257,  {\tt hep-ph/9804398}.

\bibitem{L} J.C.~Long, H.W.~Chan and J.C.~Price, Nucl. Phys. {\bf B539}
(1999) 23, {\tt hep-ph/9805217}.  

\bibitem{cvetic} 
M.~Cvetic and H.H.~Soleng,
%``Supergravity domain walls,''
Phys.\ Rept.\ {\bf 282}, 159 (1997), 
hep-th/9604090 and references therein.
%%CITATION = PRPLC,282,159;%%

\bibitem{hw}
P.~Horava and E.~Witten,
Nucl. Phys. {\bf B460} (1996) 506, 
{\tt hep-th/9510209}; 
%%CITATION = NUPHA,B460,506;%%
Nucl. Phys. {\bf B475} (1996) 94, 
{\tt hep-th/9603142}.
%%CITATION = NUPHA,B475,94;%%

\bibitem{stelle}
A.~Lukas, B.A.~Ovrut, K.S.~Stelle and D.~Waldram,
Phys. Rev. {\bf D59} (1999) 086001, 
{\tt hep-th/9803235}.
%%CITATION = PHRVA,D59,086001;%%

\bibitem{verlinde}
H.~Verlinde,
{\it Holography and compactification}, 
{\tt hep-th/9906182}.
%%CITATION = HEP-TH 9906182;%%

\bibitem{malda}
J. Maldacena,
Adv. Theor. Math. Phys. {\bf 2} (1998) 231,
{\tt hep-th/9711200};
%%CITATION = 00203,2,231;%%
S.S. Gubser, I.R. Klebanov and A.M. Polyakov,
%``Gauge theory correlators from noncritical string theory,"
Phys. Lett. {\bf B428} (1998) 105, {\tt hep-th/9802109};
%%CITATION = PHLTA,B428,105;%%
E. Witten,
%``Anti-de Sitter space and holography,"
Adv. Theor. Math. Phys. {\bf 2} (1998) 253,
{\tt hep-th/9802150}.
%%CITATION = 00203,2,253;%%

\bibitem{kehagias}
A.~Kehagias,
{\it Exponential and power law hierarchies from supergravity}, 
{\tt hep-th/9906204}.
%%CITATION = HEP-TH 9906204;%%

\bibitem{gausug}
M. Gunaydin, L.J. Romans and N.P. Warner, 
Phys. Lett. {\bf B154} (1985) 268 and 
Nucl. Phys. {\bf B272} (1985) 598; M. Pernici, 
K. Pilch and P. van Nieuwenhuizen, Nucl. Phys. {\bf B259} (1985) 460.

\bibitem{KLT}
{P. Kraus, F. Larsen and S.P. Trivedi, 
%{\it The Coulomb Branch of Gauge Theory from Rotating Branes}, 
JHEP {\bf 03} (1999) 003, {\tt hep-th/9811120}.}

\bibitem{Sfe1}
K. Sfetsos,
%``Branes for Higgs phases and exact conformal field theories,"
JHEP {\bf 01} (1999) 015, {\tt hep-th/9811167}.
%%CITATION = JHEPA,9901,015;%%

\bibitem{FGPW}
D.Z.~Freedman, S.S.~Gubser, K.~Pilch and N.P.~Warner,
{\it Continuous distributions of D3-branes and gauged supergravity},
{\tt hep-th/9906194}.
%%CITATION = HEP-TH 9906194;%%

\bibitem{BS}
A.~Brandhuber and K.~Sfetsos,
{\it Wilson loops from multicentre and rotating branes, mass gaps and phase 
structure in gauge theories}, {\tt hep-th/9906201}.
%%CITATION = HEP-TH 9906201;%%

\bibitem{CheGid}
I.~Chepelev and R.~Roiban,
{\it A note on correlation functions in AdS(5) / SYM(4) 
correspondence on the Coulomb branch}, {\tt hep-th/9906224};
%%CITATION = HEP-TH 9906224;%%
S.B. Giddings and  S.F. Ross, {\it D3-brane shells to black branes on the
Coulomb branch}, {\tt hep-th/9907204}.
%%CITATION = HEP-TH 9907204;%%

\bibitem{kesf}
A.~Kehagias and K.~Sfetsos,
{\it Deviations from the $1/r^2$ Newton law due to extra dimensions 
and higher-derivative terms}, {\tt hep-ph/9905417}.
%%CITATION = HEP-PH 9905417;%%

\bibitem{wise2}
W.D.~Goldberger and M.B.~Wise,
{\it Modulus stabilization with bulk fields}, {\tt hep-ph/9907447}.
%%CITATION = HEP-PH 9907447;%%

\bibitem{FL}
M. Floratos and G. Leontaris,
{\it Low scale unification, Newton's law and extra dimensions},
{\tt hep-ph/9906238}. 
%%CITATION = HEP-PH 9906238;%%

\bibitem{HDDK}  N. Arkani-Hamed, S. Dimopoulos, G. Dvali and N. Kaloper,  
{\it Infinitely large new dimensions},  {\tt hep-ph/9907209}.

\end{thebibliography}
\end{document}